\documentclass[twocolumn,showpacs,preprintnumbers,amsmath,amssymb,aps]{revtex4}

\usepackage{chemarrow}
\usepackage[pdftex]{graphics}  
\usepackage{graphicx}
\begin{document}


\title{Noise and critical phenomena in biochemical signaling cycles\\ at small molecule numbers}

\author{C. Metzner} \email{claus.metzner@gmx.net}
\author{M. Sajitz-Hermstein}
\author{M. Schmidberger}
\author{B. Fabry}

\affiliation{
Biophysics Group, Department of Physics, University of Erlangen, Germany
}

\date{\today}


\begin{abstract}


Biochemical reaction networks in living cells usually involve reversible covalent modification of signaling molecules, such as protein phosphorylation. Under conditions of small molecule numbers, as is frequently the case in living cells, mass action theory fails to describe the dynamics of such systems. Instead, the biochemical reactions must be treated as stochastic processes that intrinsically generate concentration fluctuations of the chemicals. We investigate the stochastic reaction kinetics of covalent modification cycles (CMCs) by analytical modeling and numerically exact Monte-Carlo simulation of the temporally fluctuating concentration. Depending on the parameter regime, we find for the probability density of the concentration qualitatively distinct classes of distribution functions, including power law distributions with a fractional and tunable exponent. These findings challenge the traditional view of biochemical control networks as deterministic computational systems and suggest that CMCs in cells can function as versatile and tunable noise generators.

\end{abstract}


\maketitle



\section{\label{sec:int}Introduction}


Living cells transduce chemical signals from the environment via trans-membrane receptors to their interior. The activated receptors trigger chains of chemical reactions along so-called signaling pathways, which can for example lead to the expression of selected genes in response to the external stimulus. Complex reaction networks arise when several linear pathways are cross-linked by multiple biochemical interactions. Such signal transduction networks are traditionally thought of as deterministic "computers", in which information is coded by the relative concentration of bio-chemicals. This study challenges this view and suggests that stochastic concentration fluctuations are the primary mode of operation for most intracellular signaling cascades.

It is well known that the numbers of receptors and signaling molecules fluctuate as a function of time and from cell to cell \cite{Fur05}. The role of these fluctuations, often regarded as noise, is still poorly understood. How can cells properly react to external stimuli when the signals have to pass through noisy channels ? Is the degree of noise actively suppressed for certain key signaling proteins ? Or is the present understanding of intra-cellular control, based on mass action theory, overly simplified ?

Using covalent modification cycles (CMCs) as a simple model system, we show that the magnitude of concentration fluctuations, relative to the mean value, can indeed be enormous. We demonstrate that CMCs can be viewed as versatile and tunable noise generators. Depending on the system parameters, qualitatively different classes of probability density functions (PDFs) of concentration fluctuations emerge, including extremely broad and asymmetric distributions with fractional power law tails. 

CMCs are a very common motif in cellular reaction networks \cite{Kit02,Kho06,Har99,Kos98,Sha84,Kre81,Sta77}. The typical structure of a CMC is shown in Fig.\ref{fig:cmc} below.
\begin{figure}[ht]
\includegraphics[width=7.0cm]{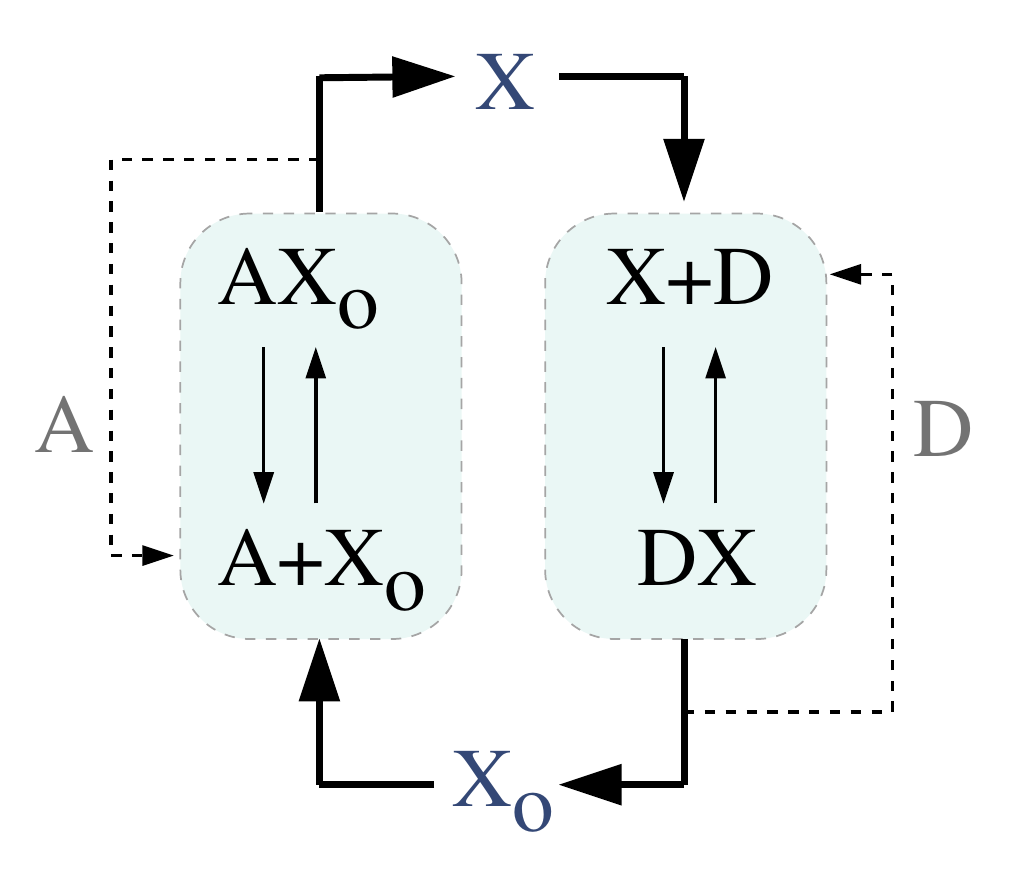}
\caption{\label{fig:cmc}{(Color online) \em Schematic of a covalent modification cycle (CMC):} Substrate $X_0$ is activated by enzyme $A$ into the modified form $X$ and deactivated by enzyme $D$. Each shaded sub-module denotes an enzymatic conversion reaction (Unbinding reactions not shown).}
\label{schema}
\end{figure}
In such systems, a substrate protein is found in two different chemical states, an inactive form $X_0$ and an activated form $X$ (often a phosphorylized version of $X_0$). The conversion of the two forms into each other is provided by an activating enzyme $A$ (often a kinase), the deactivation by another enzyme $D$ (often a phosphatase). In the activation process, the catalyst $A$ first binds its substrate $X_0$. The resulting enzyme-substrate complex $AX_0$ may decay back into the original components. In the case of a successful conversion, however, a product molecule $X$ is released and the enzyme $A$ is recovered for further use. The deactivation process is analogous. 

The CMC can be functionally decomposed into two enzymatic conversion processes. According to Michaelis-Menten kinetics (comp. Appendix \ref{prod}), the conversion rate is, in the linear regime, limited by the amount of available substrate. For very high substrate concentration, however, the conversion rate approaches a maximum value, determined only by the amount and efficiency of the enzyme (saturation regime).

As demonstrated in a classical paper by Goldbeter and Koshland \cite{Gol81}, the combination of the two enzymatic conversion reactions can lead to interesting behavior if they operate within the saturated regime. In this case, the equilibrium ratio $[X]/[X_0]$ as a function of the ratio of enzyme levels $[A]/[D]$ develops a sigmoidal shape with a sharp transition point (zero-order ultra-sensitivity). In the context of biochemical signal networks, CMCs are for this reason understood as switches. 

The Goldbeter-Koshland theory is based on deterministic (mass action) rate equations and thus disregards fluctuations entirely. Molecular reactions, however, inevitably generate intrinsic noise, due to their discrete and stochastic nature. Even under so-called steady-state conditions, the momentary rates at which reactions proceed are fluctuating around the mean values described by mass action theory. The corresponding temporal fluctuations of molecule numbers are particularly important in living cells, where the average molecule numbers of many chemical species are low. For this reason, quantitative models of biochemical concentration fluctuations are developed for different types of reaction networks (see, for example, Refs.\cite{Nac06,War06,Qia02}).

Due to their ubiquity in living cells, CMCs are of particular interest. A detailed theoretical investigation of the intrinsic fluctuations of CMCs, their robustness and tunability was provided by Levine et al.\cite{Lev07}, who directly solved the master equation for the probability distribution of the number of activated signal molecules. 
The authors further consider the information transmission properties of the system in the presence of the intrinsic fluctuations, by applying a pulse-like increase of the kinase activity as an input signal. They find that the noisy CMC can transmit the signal reliably if tuned to an optimal parameter range.

In this paper, we focus on the shape of the stationary probability distributions produced by CMCs in various parameter regimes. The reaction kinetics of this system is simulated using the exact Gillespie algorithm. This simulation yields directly the temporal concentration fluctuations $x(t)$ of the activated signaling molecule. 

We find an unexpected variety of distribution functions $P(x)$, including Gaussian, exponential, flat, as well as power law distributions with a fractional and tunable exponent. The type of the emerging distribution function depends on parameters such as the total amount of available enzyme and substrate molecules in their different forms and on reaction rate coefficients. We speculate that living cells could switch between distinct statistical distributions, on short time scales, by controlling the overall expression levels of these molecules. In many cases, moreover, the enzymes of a CMC are themselves activated and deactivated by another cycle. In this way, the effective conversion efficiency of an enzyme can be tuned over a wide range with only minimal changes of protein expression levels. This tremendous flexibility of CMCs with respect to their statistical properties suggests a more complex picture of cellular signal processing which is based on the active generation and precise shaping of concentration fluctuations of signaling molecules. 

In our paper we develop analytical approximations of the concentration fluctuations within CMCs, based on stochastic differential equations and explicit stationary solutions of the corresponding Fokker-Planck equations. The analytical results are in excellent agreement with the simulations and provide a quantitative understanding of the major statistical features.


\section{\label{sec:mod}Models and Methods}


\subsection{Model parameters and assumptions}

Let the reactions take place in a container of volume $V$, so that the concentration $[S]$ of a substance corresponds to a molecule number $s=[S]V$. We also assume that the reactor is "well-stirred", i.e. diffusion of chemical species is infinitely fast and so spatial effects are disregarded.

We study a CMC of the form
\begin{eqnarray}
X_0 + A &\autorightleftharpoons{$b_1$}{$u_1$}& AX_0 \autorightarrow{$c_1$}{} X + A\nonumber\\
X + D &\autorightleftharpoons{$b_2$}{$u_2$}& DX \autorightarrow{$c_2$}{} X_0 + D.
\label{chemeq}
\end{eqnarray}
The substrate $X_0$ is converted into its activated form $X$ by enzyme $A$. The corresponding deactivation is performed by enzyme $D$. We thus have to consider 6 temporally variable molecule numbers $ x_0, x, a, d, ax_0, dx $, dynamically coupled by 6 chemical reactions. Within each enzymatic conversion unit, the 3 reaction coefficients are denoted $b$ (binding), $u$ (unbinding) and $c$ (conversion). Index 1 is used for the activation and index 2 for the deactivation process. Additional parameters are the total amount of the substrate in its various forms, $ x_t = x_0 + ax_0 + x + dx $, as well as the total amounts of enzymes $ a_t=a+ax_0 $ and $ d_t=d+dx $.


\subsection{Analytical and numerical methods}

Analytical approximations for chemical reaction networks can be obtained by deriving Langevin equations for the temporal changes of the molecule numbers. These stochastic differential equations contain, besides a deterministic term that corresponds to the mass action change rates, a stochastic term that accounts for the fluctuations. To make use of the standard methods of stochastic differential calculus, the fluctuation term is approximated by a Gaussian, white noise random process. This is a critical approximation, since the effective "strength" of the white noise process has to be chosen with care, in order to reflect the true process as faithfully as possible. In the case of chemical Langevin equations, the true process consists of a series of delta-peaks, arriving with (inhomogeneous) Poisson statistics. It is therefore possible to derive the proper strength of the white noise process from the fundamental properties of Poisson statistics. This theory of chemical Langevin equations has been developed, for the general case, by Gillespie \cite{Gil00}. In this paper, we take a similar approach, suitable for our specific reaction network.

In order to test our analytical approximations, we shall compare the results with a numerically exact Monte-Carlo-Simulation of the reaction dynamics
by implementing the Gillespie algorithm \cite{Gil77}.
In this algorithm, the molecule numbers of each species
are integers which change abruptly due to elementary reaction
events. Statistically, these elementary reactions
are Poisson-processes with average event rates depending
on the momentary molecule numbers, according to
the chemical rate equations. Therefore, the intrinsic
stochastic fluctuations of the reactions are automatically
included in a realistic way.


\subsection{Coarse graining of the enzymatic conversion}

We first focus on a single enzymatic conversion reaction, for example the activation process. Our goal is to describe it in a coarse grained approximation as a single functional unit with effective statistical properties. Two of these effective units will later be combined (as shown in Fig.\ref{fig:cmc}) to derive a stochastic differential equation for the fluctuating number $ x(t) $ of $X$-molecules.

We assume for a moment that the number $ x_0 $ of substrate molecules $X_0$ is constant (ideal reservoir). We are then interested in the average production rate $ \overline{R}_{act}(x_0) $ of the activated protein $X$ and in the temporal fluctuations $ \Delta R_{act}(x_0,t) $ of this rate. This, in turn, will enable us to write a stochastic rate equation of the production process in the form $ \dot{x}=\overline{R}_{act}(x_0)+\Delta R_{act}(x_0,t) $.

As for the average rates, we solve the mass action rate equations in the stationary state. This follows standard Michaelis-Menten theory, but for completeness we include the derivation in Appendix \ref{prod}. The result is 

\begin{equation}
\overline{R}_{act}(x_0) = v_m \frac{x_0}{x_0+k_m}
\end{equation}

with the maximum conversion velocity

\begin{equation}
v_m=c a_t
\end{equation}

and the Michaelis constant

\begin{equation}
k_m=\frac{c+u}{b}.\label{mmc}
\end{equation}

Enzymes are sometimes likened to nano machines, which convert their substrates in a predictable, goal-oriented process. Yet, many enzymes in biological systems are working in a much more imperfect way: Once the enzyme has bound to its substrate, the enzyme-substrate-complex often dissociates back into the original two molecules. Each individual enzyme molecule will go through a series of futile binding-dissociation cycles, before it actually converts a substrate into the modified form. In the chemical reaction equation (\ref{chemeq}), this is accounted for by the back reaction with rates $u_j$ (with $j=1,2$). The conversion efficiency of an enzyme can be quantified by the fraction of binding events that lead to a successful production and release of the modified substrate molecule. This fraction, in turn, depends on the relative magnitude of the rates $u_j$ and $c_j$.
We can define two limiting regimes: The case $u_j>>c_j$ corresponds to extremely inefficient enzymes. In the diagram of Fig.(\ref{schema}), almost all activity of the reaction system will then take place within the shaded sub-modules. The flux in and out of these sub-modules is so weak that within the sub-modules a chemical equilibrium is established between the bound and dissociated enzyme-substrate-complexes. We therefore call this case the "pre-equilibrium" regime. The opposite case, $c_j>>u_j$ corresponds to highly efficient enzymes. In the diagram of Fig.(\ref{schema}), the system is running uni-directionally around the cycle, for most of the time. We therefore call this case the "sequential" regime.


Independently from $u$ and $c$, two other limiting regimes are connected with the amount of substrate $x_0$ relative to the Michaelis constant $k_m$. The system is in the `linear' regime for $x_0<<k_m$ and in the `saturation' regime for $x_0>>k_m$.

Next we model the fluctuations $ \Delta R_{act}(x_0,t) $ of the production rate around the average value $ \overline{R}_{act}(x_0)$. The statistical properties of these fluctuations are not obvious, even if the substrate molecule number $ x_0 $ is artificially held constant. As motivated in Appendix \ref{poiss}, we approximate the production process, in a coarse grained view, as a Poisson process with average event rate $ \overline{R}_{act}(x_0) $. Numerical simulations, shown below, confirm that the probability distribution of the waiting time between successive $X$-production events is indeed exponentially distributed with the expected characteristic time constant. We further approximate the above Poisson process by white Gaussian noise with a proper pre\-fac\-tor(Appendix \ref{gauss}). As a result of the above coarse-graining procedure, we obtain
\begin{equation}
\dot{x} = \overline{R}_{act}(x_0) + \sqrt{\overline{R}_{act}(x_0)}\cdot \zeta(t),
\end{equation}
where $ \zeta(t) $ is normalized white Gaussian noise with $ \left\langle \zeta(t)\zeta(t^{\prime})\right\rangle = \delta(t-t^{\prime}) $. 

\subsection{Stochastic differential equation of a CMC}

We next combine the activation and deactivation processes. The molecule numbers $x(t)$ and $x_0(t)$ are now both considered as variables. One obtains 

\begin{eqnarray}
\dot{x} &=& \left[ \; \overline{R}_{act}(x_0) - \overline{R}_{dea}(x) \right]\nonumber\\ 
&+& \left[ \sqrt{\overline{R}_{act}(x_0)}\cdot \zeta_a(t) + \sqrt{\overline{R}_{dea}(x)}\cdot \zeta_d(t) \right].
\end{eqnarray}

Note that the deactivation rates depend on $x$, not $x_0$. To make further progress, we neglect the amount of substrates bound within enzyme-substrate complexes, so that $ x_0=x_t-x $. Additionally, we make the simplifying assumption that the noise terms of the activation and deactivation processes fluctuate statistically independent from each other. We can then combine both terms, adding up the variances:

\begin{eqnarray}
\dot{x} &=& \left[ \; \overline{R}_{act}(x_t-x) - \overline{R}_{dea}(x) \right]\nonumber\\ 
&+& \left[ \sqrt{\overline{R}_{act}(x_t-x) + \overline{R}_{dea}(x)} \right]\cdot \zeta(t).
\end{eqnarray}

This has the general form of a stochastic differential equation with a multiplicative noise term \footnote{Note that stochastic differential equations of the general form $\dot{x} = f(x) + g(x)\cdot \zeta(t)$ are extremely rich in behavior and can produce random fluctuations with arbitrary PDF and ACF, as shown in Ref. \cite{Pri99} and \cite{Pri00}}:

\begin{equation}
\dot{x} = f(x) + g(x)\cdot \zeta(t).
\label{sde}
\end{equation}

Here,

\begin{equation}
f(x)=v_a \frac{(x_t-x)}{(x_t-x)+k_a} - v_d \frac{x}{x+k_d}
\label{fff}
\end{equation}

and

\begin{equation}
g(x) = \sqrt{\; v_a \frac{(x_t-x)}{(x_t-x)+k_a} +  v_d \frac{x}{x+k_d}  },
\label{ggg}
\end{equation}

with obvious definitions of $ v_a, v_d, k_a, k_d $. In the following, we will extract statistical properties of this random process. Note that the Ito interpretation has to be used, whenever the true random process (that is to be approximated by Gaussian white noise) consists of a series of $\delta-$peaks, such as in our case of intrinsic, chemical noise \cite{Ris84,Kam92}. 

We define a drift term,

\begin{equation}
A(x)=f(x)
\label{driftterm}
\end{equation}

and a diffusion term

\begin{equation}
B(x)= \frac{1}{2}g^2(x).
\end{equation}

The time-dependent PDF $ P(x,t) $ of the fluctuating variable x(t) approximately satisfies the 
Fokker-Planck equation

\begin{equation}
\frac{\partial}{\partial t}P(x,t)=-\frac{\partial}{\partial x}\left[ A(x)P(x,t) \right] + \frac{\partial^2}{\partial x^2}\left[ B(x)P(x,t) \right].
\end{equation}

The stationary solution $ P(x) $ of this equation reads

\begin{equation}
P(x)=\frac{N}{B(x)} \exp\left[ \int_{x_{min}}^x \!\!\frac{A(s)}{B(s)}ds \right].
\end{equation}

Here, $N$ is a normalization constant.


\subsection{The symmetric CMC \label{symcmc}}

With $ v_a, v_d, k_a, k_d $ and $ x_t $, there is obviously a large parameter space to explore. In this paper, we shall restrict ourselves to just a few interesting cases. In a symmetric CMC, the activation and deactivation processes have the same parameters, i.e. $ v_a=v_d=v $ and $ k_a=k_d=k $. We then have

\begin{equation}
f(x)=v\left[\; \frac{(x_t-x)}{(x_t-x)+k} - \frac{x}{x+k} \right]
\label{fff}
\end{equation}

and

\begin{equation}
g^2(x) = v \left[ \frac{(x_t-x)}{(x_t-x)+k} + \frac{x}{x+k} \right].
\label{ggg}
\end{equation}

Because the drift term $A(s)$ and the diffusion term $B(s)$ are both proportional to $v$, it is clear that the maximum production rate $v$ will not affect the shape of the stationary PDF. Consequently, $ k $ and $ x_t $ are the only important parameters left.

\subsubsection{Linear Regime}

The limit of a large Michaelis constant, $ k>>x_t $, corresponds to the linear regime of the two enzymatic conversion reactions. In this case, the terms $x$ and $(x_t-x)$ can be neglected in Eqs. (\ref{fff}) and (\ref{ggg}).
This leaves us with 
\begin{equation}
f(x)=(v x_t / k)- (2v/k)x
\end{equation}
and 
\begin{equation}
g^2(x) = (v x_t / k).
\end{equation}
A straight forward calculation of the PDF yields a Gaussian, centered at $ \overline{x}=\frac{x_t}{2} $, with a variance $ \sigma_x^2=\frac{x_t}{4} $:

\begin{equation}
P(x) \propto e^{-\frac{2(x-(x_t/2))^2}{x_t}}
\end{equation}

The stochastic differential equation of a symmetric, linear CMC corresponds to an Ornstein-Uhlenbeck process. Besides the Gaussian PDF, we therefore expect an exponentially decaying autocorrelation function:
\begin{equation}
C_{xx}(\tau)=<\Delta x(\tau) \Delta x(0) > = \left( \frac{x_t}{4} \right)\; e^{-(2v/k) \tau}.
\end{equation}
The characteristic time constant is $\tau_c=\frac{k}{2v}$.

\subsubsection{Saturation Regime}

Next, we consider the opposite case of a small Michaelis constant, i.e. $ k<<x_t $, corresponding to the saturation regime. We then have
\begin{equation}
f(x)=v\left[ 1-\frac{x}{x+k} \right]\;\rightarrow\; \frac{vk}{x}\;\;\mbox{for}\;\;x>>k
\end{equation}
and 
\begin{equation}
g^2(x) = v\left[ 1+\frac{x}{x+k} \right]\;\rightarrow\; 2v\;\;\mbox{for}\;\;x>>k.
\end{equation}

The asymptotic drift and diffusion terms are $ A(x)= \frac{vk}{x}$ , $ B(x)=v $, and $ A(s)/B(s)=\frac{k}{x} $. Therefore, 
\begin{equation}
\int_{x_{min}}^x \; \frac{A(s)}{B(s)}ds = k \cdot \log(x/x_{min}),
\end{equation}
and
\begin{equation}
P(x) \propto e^{k \cdot \log(x/x_{min})} \propto (x/x_{min})^k.
\end{equation}
Hence, we expect an increasing power-law tail for the asymptotic PDF in the saturation regime of the symmetric CMC. The exponent of the power-law can be fractional and is equal to the dimensionless Michaelis constant (Eq.\ref{mmc}). The above analytical approximations will break down when $x$ approaches the limits $ 0 $ or $ x_t $.



\subsection{\label{sec:asyPar}The asymmetric CMC}

We now allow the activation parameters $k_a$ and $v_a$ to differ from the corresponding deactivation parameters $k_d$ and $v_d$. Under saturation conditions ($x_t>>k_a$, $x_t>>k_d$) and in the limit of large $x$ one obtains
$f(x)\rightarrow(v_a-v_d)$ and $g^2(x)\rightarrow(v_a+v_d)$, so that 
\begin{equation}
\frac{A(s)}{B(s)}\rightarrow \;\lambda\; = 2\frac{v_a-v_d}{v_a+v_d}.
\end{equation}
This results in a stationary PDF with an exponential tail:
\begin{equation}
P(x) \propto e^{\lambda x}.
\end{equation}
The decay constant $\lambda$ is positive for $v_a>v_d$ and negative for $v_a<v_d$.


\section{\label{sec:res}Results}



\subsection{Validation of Poisson statistics}

We first investigate the statistics of the enzymatic activation process, with artificially fixed number $x_0$ of substrate molecules. For this purpose, we perform direct Monte-Carlo simulations in different parameter regimes. All rates and times are presented in dimensionless numbers.

\begin{figure}[ht]
\includegraphics[width=8.0cm]{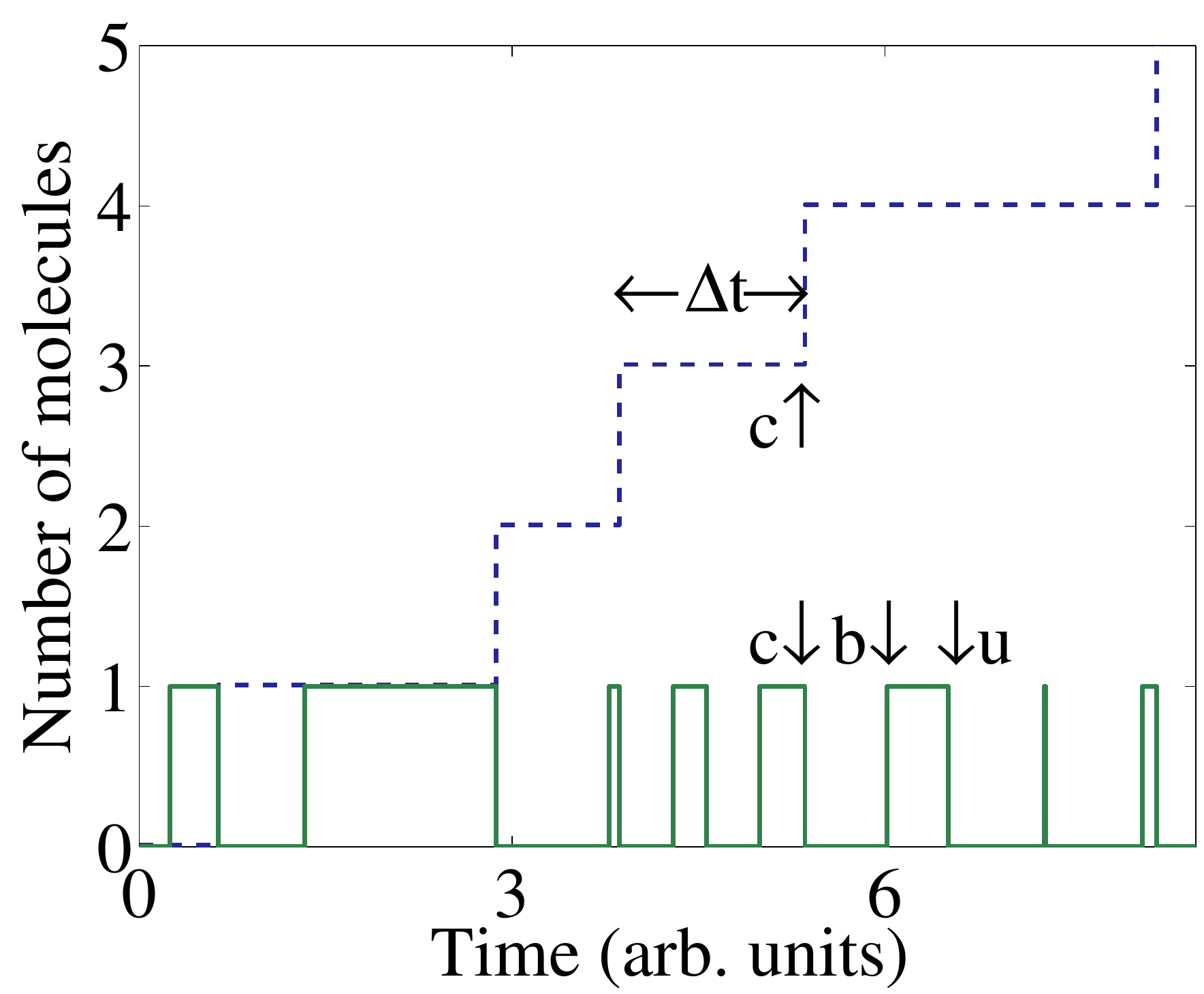}
\caption{\label{fig:idealMM} {(Color online) \em Monte-Carlo simulation of enzymatic conversion:} Molecule numbers of the enzyme-substrate complex (solid) and of the activated product (dashed) in the case of only one enzyme molecule. Parameters: $b=u=c=1.0$, $e_t=1$. The vertical arrows denote a conversion ($c$), binding ($b$) and unbinding ($u$) process. $\Delta t$ is the time interval between two successive conversion events.}
\end{figure}

The stochastic time evolution of the enzymatic activation process is characterized by abrupt changes of the various molecule numbers by integer amounts (Fig. \ref{fig:idealMM}). A single enzyme molecule sometimes undergoes binding ($b$) and unbinding ($u$) without conversion ($c$) to a product molecule. The time interval $ \Delta t $ between two successive conversion events is fluctuating around the inverse of the average production rate. 


\begin{figure}[ht]
\includegraphics[width=8.0cm]{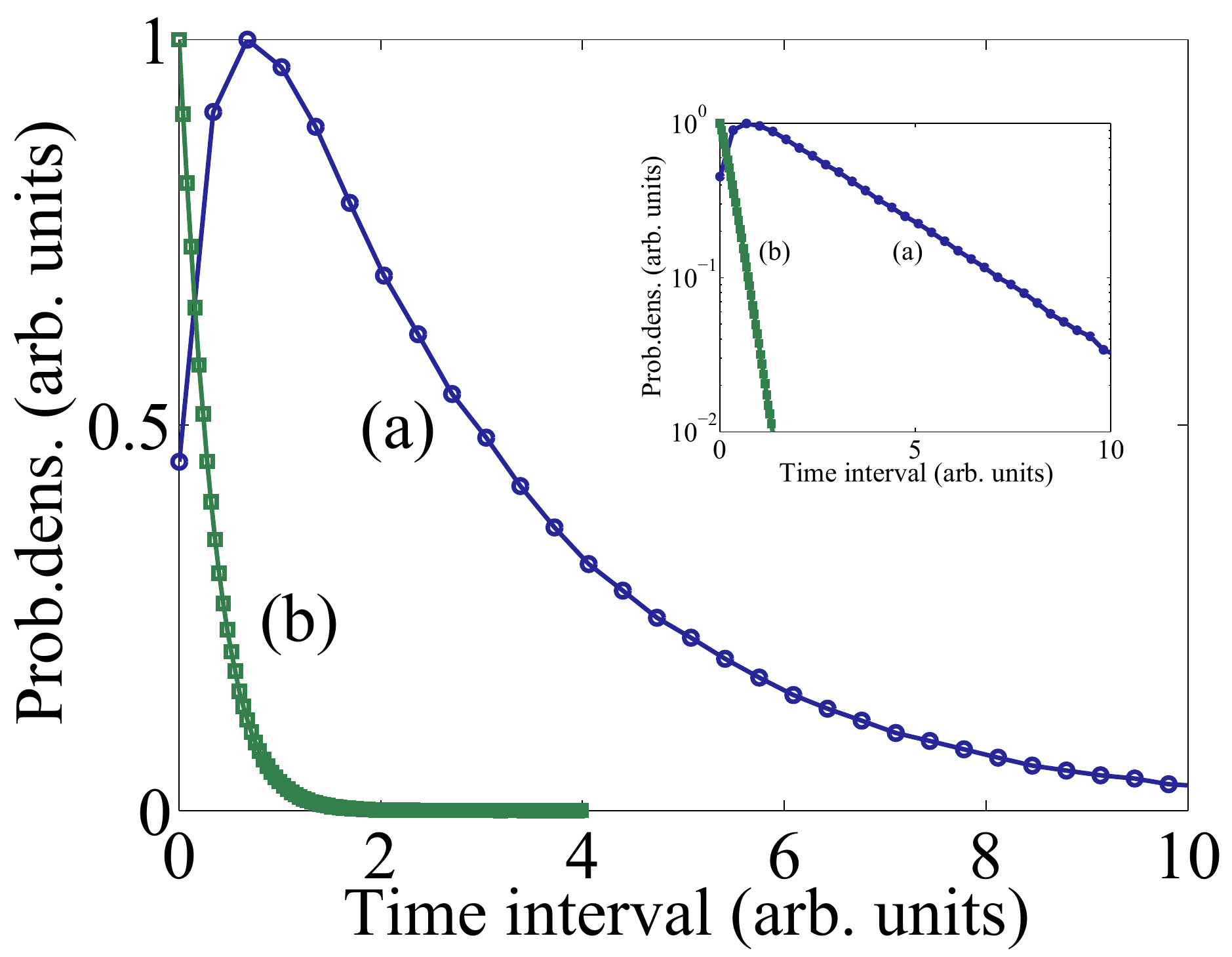}
\caption{\label{fig:dtStat}{(Color online)\em  Waiting time distributions:} Simulated PDF of the time intervals between successive conversion events. Parameters: $b=u=c=1.0$, $x_0=1=const.$. Case (a): Only one enzyme molecule. Case (b): 10 independent enzyme molecules. The inset shows the same data in a semi-logarithmic plot.}
\end{figure}

Since the production process involves a sequence of elementary reaction steps, the distribution function $P(\Delta t)$ of this waiting time is not expected to be exponential for an individual enzyme molecule. However, the superposition of many such multi-step processes running independently from each other can closely mimic a Poisson process (Fig.\ref{fig:dtStat}).



\subsection{Monte-Carlo Simulation of the CMC}

Next we discuss the statistical properties of CMCs in selected parameter regimes, as obtained by Monte-Carlo simulation of the reaction dynamics. We shall mainly focus on CMCs with symmetric parameters for the activation and deactivation process. The total number of substrate molecules $x_t$ was 200 in all cases. Our analytic theory was based on the assumption that the amount of substrate bound in complexes is small compared to $x_t$. We have therefore chosen a small number of enzyme molecules, $a_t\!=\!d_t\!=\!$10. The (rounded) parameters for all following simulations are listed in Tab.(\ref{partab}). We have also included a saturation parameter (SP), defined as $SP=x_t/k_m$, and an equilibrium parameter (EP), defined as $EP=c/u$. For instance, $SP\!\gg\! 1, EP\!\gg\! 1$, would indicate that the system is in the saturated, pre-equilibrium regime.

\begin{table}[h]
\begin{center}

\begin{tabular}{|c |c| c| c| c| l| c| c| c| } \hline

fig. & subf. & $b$ & $u$ & $c$ & asym. & $k_m$ & SP & EP \\ \hline\hline
 \ref{fig:Sub_LinAndWSa},\ref{fig:Enz_LinAndWSa}   & b & $5\! \cdot10^{-4}$ & $0.1$ & $10$ & - & $2\!\cdot10^4$ & 0.01 & 0.01 \\ \hline
    & c & $5\! \cdot10^{-4}$ & $10$ & $0.1$ & - & $2\!\cdot10^4$ & 0.01 & 100 \\ \hline
    & d & $0.5$ & $0.1$ & $10$ & - & 20 & 10 & 0.01 \\ \hline
    & e & $0.5$ & $10$ & $0.1$ & - & 20 & 10 & 100 \\ \hline\hline

\ref{fig:SymPLPDF} & a & 1& 0.1& 2& - & 2.1 & 95 & 0.05\\ \hline
  & b & 1& 0.1& 1.5 & - & 1.6 & 125 & 0.07 \\ \hline
  & c & 1& 0.1& 1 & - &1.1 & 180 & 0.1 \\ \hline
  & d & 1& 0.1& 0.4 & - &0.5 & 400 & 0.25\\ \hline
  & e & 1& 0.1&  0.1& - & 0.2& 1000& 1\\ \hline\hline

\ref{fig:AsymPDF} & a & 1& 0.1&1 & $c_1\!=\!1.5$ &- &- &-\\ \hline
 & b & 1&0.1 & 1& $c_1\!=\!1.25$ &- &- &-\\ \hline
 & c & 1& 0.1&1 & $c_1\!=\!1.125$ &- &- &-\\ \hline
 & d & 1& 0.1&1 & -& 1.1&180 &0.1\\ \hline
 
\end{tabular}

\caption{\label{partab}. Parameter space explored in Monte-Carlo simulations.}
\end{center}
\vspace{-0.6cm}
\end{table}

\subsubsection{Symmetric CMC in the linear and weakly saturated regimes}

\begin{figure}[ht] 
\includegraphics[width=8.0cm]{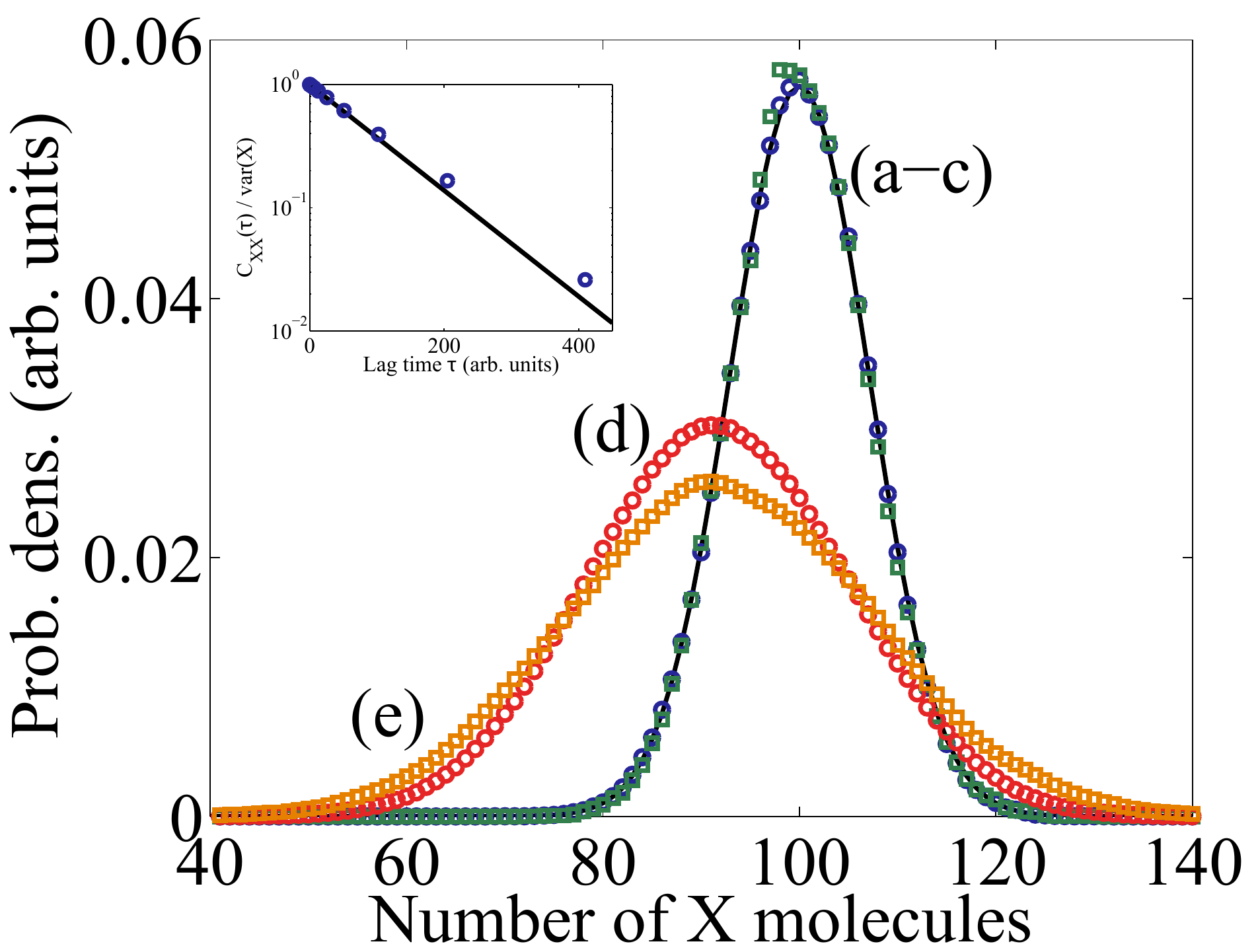}
\caption{
\label{fig:Sub_LinAndWSa}
{(Color online) \em X-distributions in the linear regime (b,c) and in the weakly saturated regime (d,e). The solid line (a) is the analytical solution to the linear case. Inset: Normalized auto-correlation function for linear case (symbols) with analytical solution (solid line). Parameters see Tab.(\ref{partab}).}
}
\end{figure}

In the linear regime, we expect for the substrate X a Gaussian distribution, peaked at $\overline{x}=x_t/2$ and with variance $x_t/4$. The autocorrelation of the random variable $x(t)$ should decay exponentially with time constant $\tau_c=k/2v$. The agreement of the Monte-Carlo results with this analytic theory is excellent (see Fig.(\ref{fig:Sub_LinAndWSa})). In the weakly saturated regime, we find a decrease of the average molecule number and a considerable broadening of the distribution, while the shape of the PDF remains approximately Gaussian. The distributions do not change dramatically when the parameter regime is changed from sequential to pre-equilibrium conditions, as long as the ratio of enzyme to substrate molecules is small (see footnote
\footnote{Note that in section \ref{symcmc} we have neglected the amount of substrate which is bound in complexes. In order to refine the theory, let us define a new dynamic variable $\alpha=x+dx$ (The complementary variable $\beta=x_0+ax_0$ is unnecessary, since $\beta=x_t-\alpha$). This variable $\alpha$ defines the macro-state of the system in our coarse-grained view. It is changed only by activation or deactivation processes. On the other hand, binding and unbinding processes only affect the micro-state of the system. The latter is defined by the numbers $dx$ and $ax_0$, each of which can vary between $0$ and the respective number of enzyme molecules. Thus, each macro state $\alpha$ can be sub-divided into several micro states $(dx,ax_0)$. The fluctuations of our variable of interest, $x(t)$, are determined by changes of the macro- and of the micro-state. In the pre-equilibrium regime, for each momentary macro-state $\alpha$, we expect that equilibrium distributions $P_{eq}(dx | \alpha)$ (and $P_{eq}(ax_0 | \beta)$) of micro-states are building up. The probability of having $x$ activated substrate molecules is under such conditions given by $P(x) = \sum_{\alpha\geq x} \;P(\alpha)\; P_{eq}(dx=\alpha-x \;|\; \alpha)$.
}).

\begin{figure}[ht]
\includegraphics[width=8.0cm]{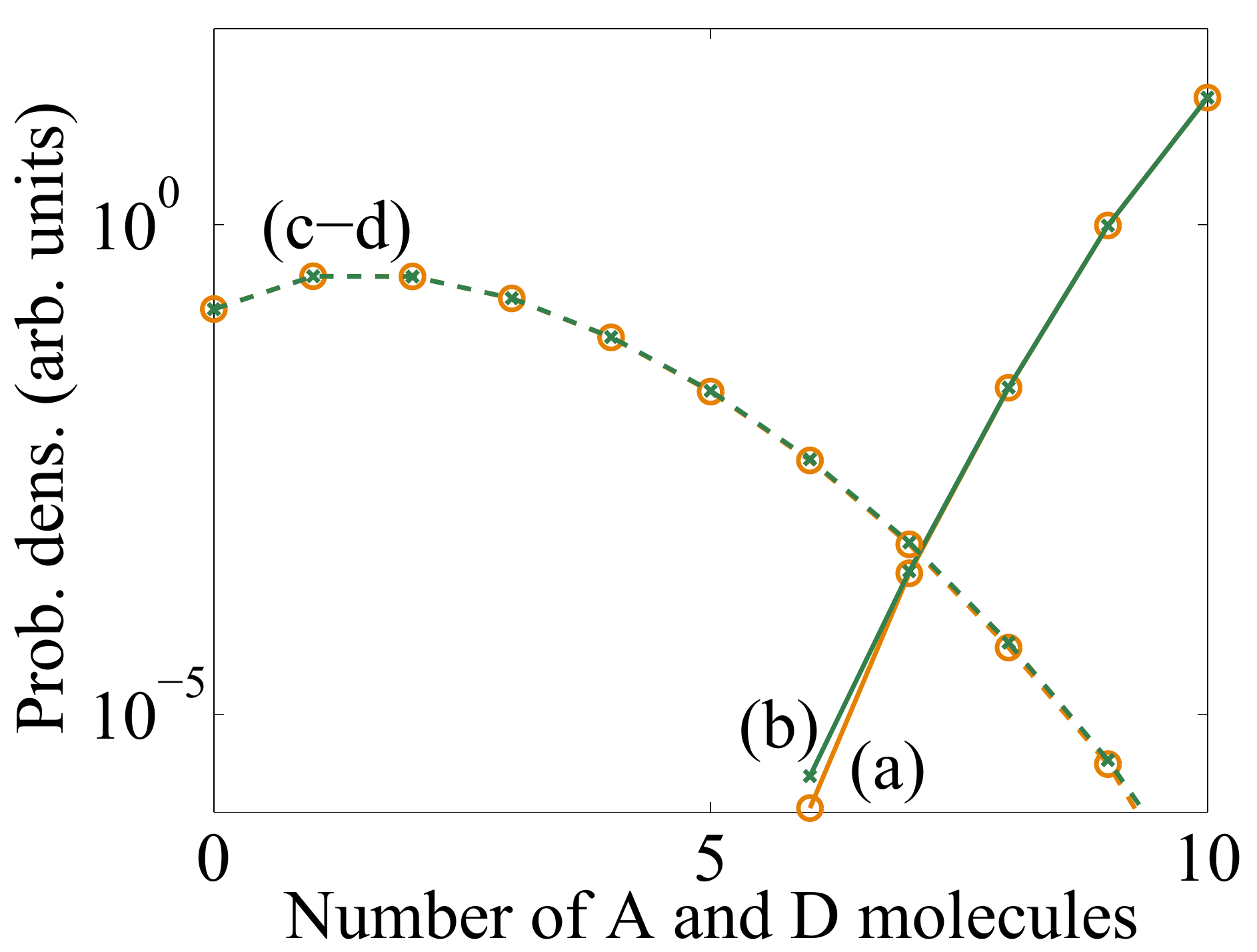}
\caption{\label{fig:Enz_LinAndWSa} 
{(Color online) \em  Semilogarithmic plots of the simulated enzyme distributions in the linear regime (a,b) and in the weakly saturated regime (c,d). Parameters see Tab.(\ref{partab}).}
}
\end{figure}

The Monte-Carlo simulations also yield the distributions of the enzyme molecule number $E\!=\!a\!=\!d$ (see Fig.(\ref{fig:Enz_LinAndWSa})). The effect of sequential or pre-equilibrium conditions is almost invisible for the particular parameters chosen.

\subsubsection{Symmetric CMC in the saturation regime}

\begin{figure}[ht]
\includegraphics[width=8.0cm]{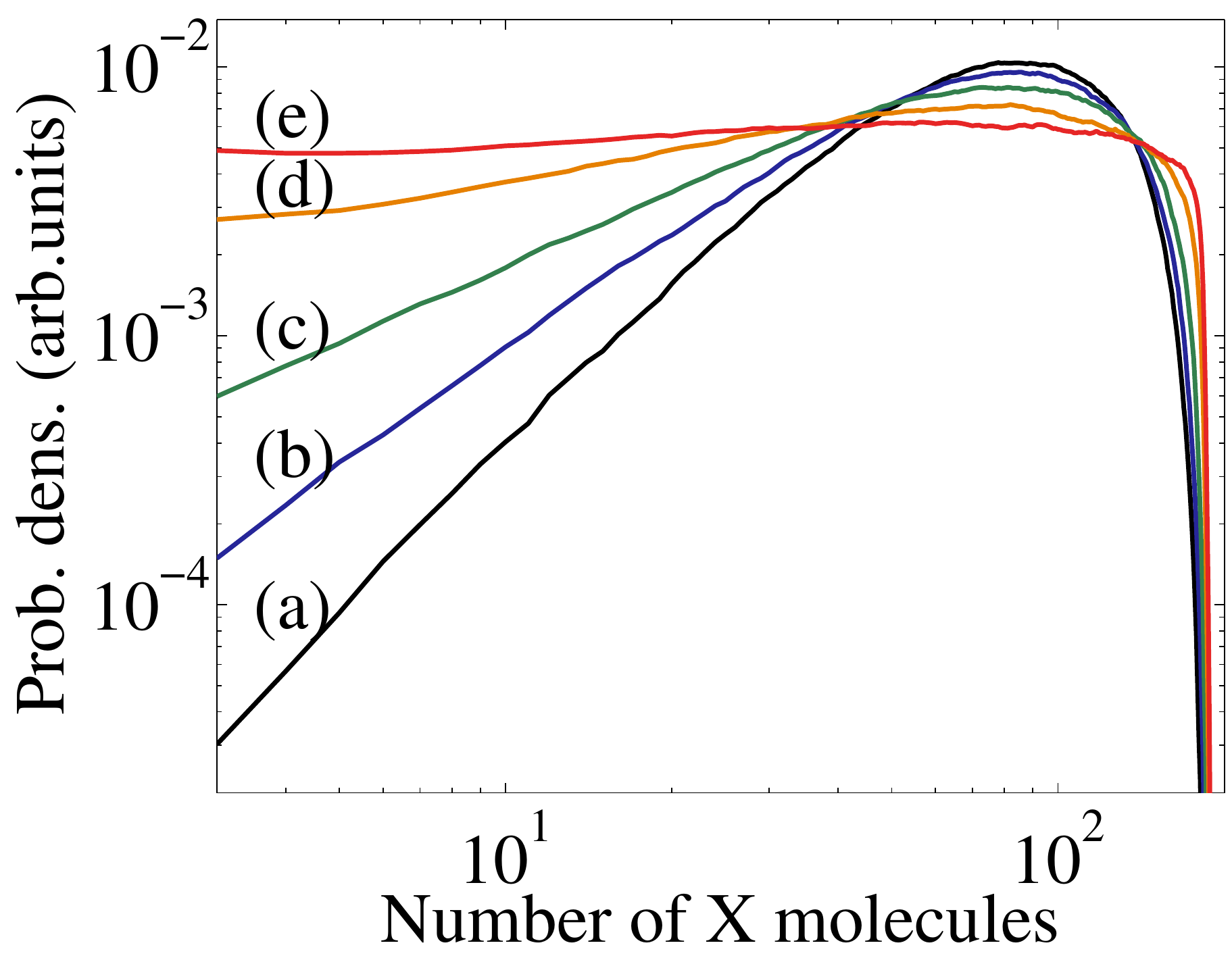}
\caption{\label{fig:SymPLPDF}{(Color online)\em  Power law tails in the strongly saturation regime:} Double-logarithmic plots of the X-distribution. In cases (a)-(e) the conversion rate $c$ has been gradually increased. Parameters see Tab.(\ref{partab}).}
\end{figure}

Next, we turn to CMCs operating within the saturation regime, which corresponds to the hypersensitive, 'switch-like' mode of the cycle. In the simulations, $k_m$ was indirectly changed via the conversion rate $c$. 
While small conversion rates result in a Gaussian PDF, the distributions become extremely asymmetric as the system runs into the saturation regime (Fig. \ref{fig:SymPLPDF}). The double-logarithmic plot reveals a power law wing at the 'left' side of the peak. The positive exponent of the power law tail is fractional in the general case. It is determined by the Michaelis constant, as expected from the analytical theory above. For a very small Michaelis constant, one obtains an almost flat distribution, which can cover several decades of concentration. Of course, the PDF has sharp cutoffs at the maximum particle number $x=x_t$ and close to $x=0$ (not shown). 

This remarkable result demonstrates that the notion of deterministic biomolecular networks, with well-defined average levels of concentration and negligibly small Gaussian fluctuations, dramatically fails in certain parameter ranges. Concentration fluctuations with a power law wing are scale-free, and therefore arbitrarily large deviations from the average value occur with non-negligible probability.

\subsubsection{Asymmetric CMC in the saturation regime}

From a systems biology point of view, an interesting question is the sensitivity of the CMC with respect to its parameter values. In particular, we investigated the effects of tuning the system slightly away from the completely symmetric parameter settings considered so far. The most dramatic effects are expected for a CMC in the hypersensitive saturation regime.

For this purpose, we start again with the parameters of the symmetric saturated CMC, which produced a PDF with a power law tail of slope 1.1 (compare Fig. \ref{fig:SymPLPDF}(e)). Now, however, we fine-tune the conversion rate $c_1$ of the activation reaction, while leaving the corresponding parameter $c_2$ at its former value $1$.

\begin{figure}[ht]
\includegraphics[width=8.0cm]{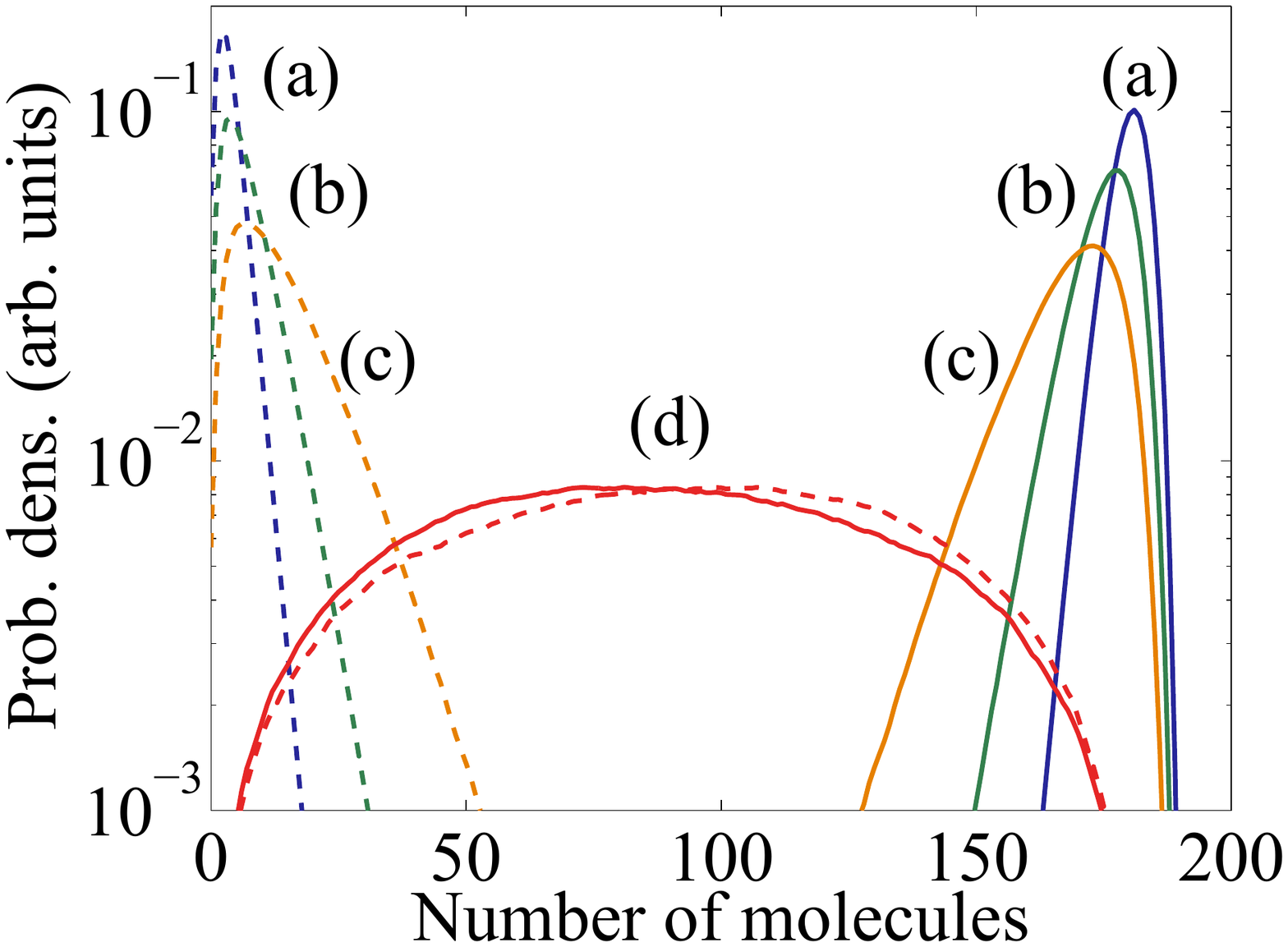}
\caption{\label{fig:AsymPDF}{(Color online)\em  Collapse of exponential distributions at the critical point of parametric symmetry:} Distributions of $X_0$ (dashed lines) and $X$ (solid lines). In cases (a)-(d) the conversion rate of the activating reaction only has been gradually increased. The left wing of (d) corresponds to Fig.\ref{fig:SymPLPDF}(c) when plotted double-logarithmically. Parameters see Tab.(\ref{partab}).}
\end{figure}

As expected, if $c_2<c_1$, the PDF of $X_0$ is peaked around a small average concentration, while $X$ has a high average concentration (Fig. \ref{fig:AsymPDF}). The average concentrations are drastically different even for rather similar $c$-parameters, due to the hypersensitive response of the saturated CMC. We find PDFs with exponential tails for all cases, except in a very narrow range around perfect parametric symmetry. This is in agreement with the analytical theory presented in section \ref{sec:asyPar}.
In the narrow symmetrical regime, the two PDFs collapse to one. They are mirror-symmetric with respect to the average molecule number in this case.

This behavior somewhat resembles critical phenomena in physics, where fluctuations of arbitrary size occur when a control parameter is precisely tuned to a critical value. 

In biological systems, it would be extremely improbable to find a CMC where all the microscopic parameters of the activation and deactivation reaction are precisely identical. However, equations (\ref{fff}) and (\ref{ggg}) show that effective dynamical symmetry can be achieved under the much weaker conditions $v_a\approx v_d$ and $k_a\approx k_d$. In terms of the microscopic parameters, this translates into $c_1 a_t\approx c_2 d_t$ and $\frac{c_1+u_1}{b_1}\approx\frac{c_2+u_2}{b_2}$. 

\begin{figure}[ht]
\includegraphics[width=7.0cm]{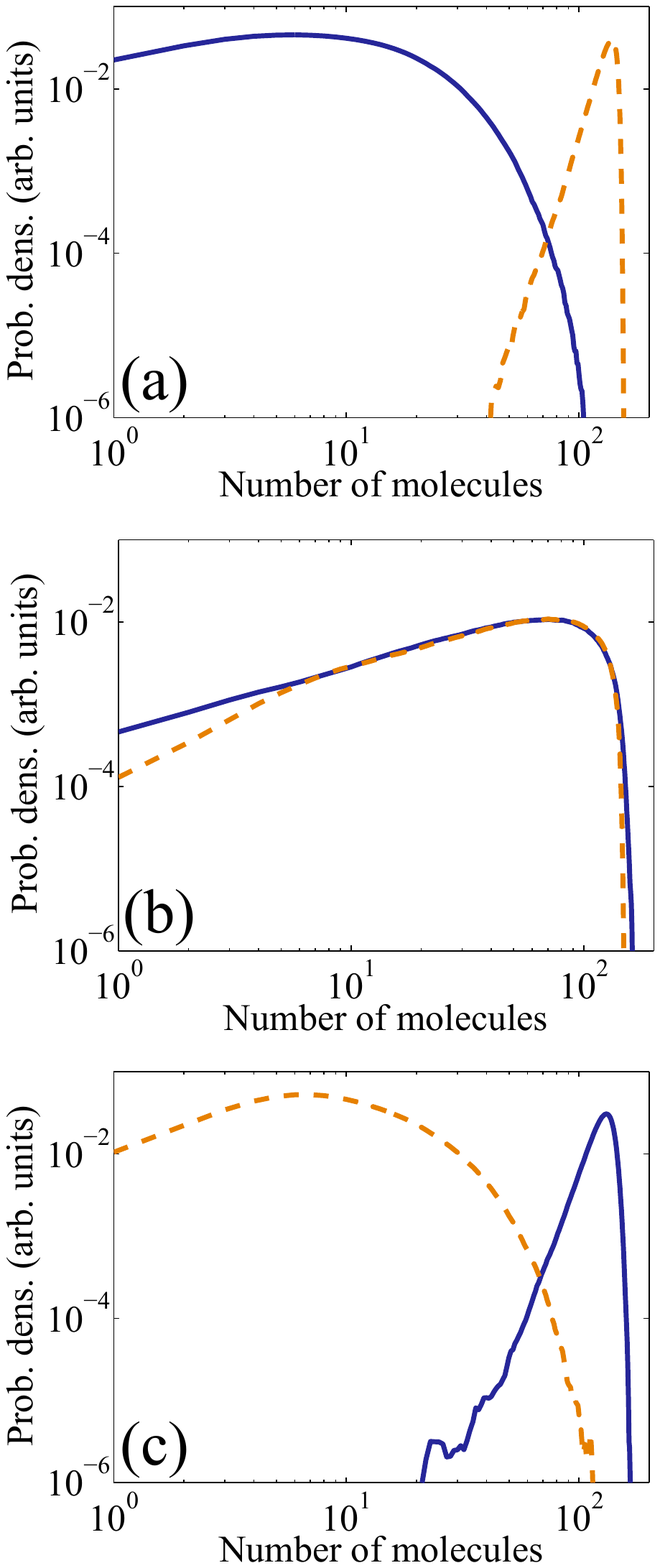}
\caption{\label{fig:Tuning}{(Color online)\em  Tuning the CMC through the critical point by changing the enzyme concentration $a_t$.} The simulated CMC has asymmetric rate constants, but becomes symmetric with respect to the effective coarse-grained parameters $k_m$ and $v_m$ for $a_t=50$. Parts (a)-(c) correspond to $a_t=$ 45, 50 and 55. Shown are double-logarithmic distributions of the activated (solid lines) and deactivated (dashed lines) substrates. For parameters see text.}
\end{figure}

In order to demonstrate that the dynamics is only controlled by the conversion velocities, the Michaelis constants and the total amount of substrate $x_t$, we have performed another Monte-Carlo simulation for a CMC with $x_t=200$ and the microscopically non-symmetric parameters $a_t\!=\!50,b_1\!=\!1,u_1\!=\!0.1,c_1\!=\!1$ for the activation reaction and $d_t\!=\!10,b_2\!=\!10,u_2\!=\!6,c_2\!=\!5$ for the deactivation reaction. These parameters are nevertheless symmetrical on the coarse-grained level of $k$ and $v$. The simulation results indeed show a power law behavior, thus confirming the analytical prediction (see Fig.(\ref{fig:Tuning})). Note that the total amount of enzymes $a_t$ and $d_t$ can be easily varied in a living cell, for example by changing the expression levels or the activity of the enzymes. This offers a way to tune the CMC through the critical point. If, for instance, we detune $a_t$ away from the critical value $a_t^{(crit)}\!=\!50$ by $\pm$10 percent, we find that one of the distributions $P(X)$ and $P(X_0)$ is loosing its power law behavior. Yet, the (respective) complementary form of substrate still shows a very steep power law tail under these conditions of disturbed symmetry.

\section{\label{sec:out}Discussion and Outlook}


The statistical properties of concentration fluctuations produced by CMCs reveal an extremely rich behavior. A variety of qualitatively different probability distributions has been found for the molecule numbers of the activated substrate, depending on the parameter settings. A particularly remarkable result for symmetric CMCs operated in the saturation regime was the emergence of an extremely broad PDF with a power law tail. These fluctuations are driven by purely intrinsic noise, originating from the stochastic arrival times of the molecular reaction events.

We note that in biological systems there are additional, extrinsic sources of noise as well. For example, we have considered the total number of enzyme molecules, $a_t$ and $d_t$, as being strictly constant in this paper. In biological systems, the enzymes are themselves subject to production and consumption processes and will therefore undergo concentration fluctuations. When these enzymes serve a CMC in the saturation regime, the steady state activation level $x/x_0$ of the substrate will depend hyper-sensitively on the momentary ratio of enzyme concentrations $a_t/d_t$. Small (and sufficiently slow) fluctuations of the enzyme concentrations will therefore be amplified, leading to an additional, extrinsic broadening of the PDF of $x(t)$. 

At first glance, it seems that such extreme concentration fluctuations would compromise the function of biochemical networks \cite{Tha01,Tha02}. However, recent reports have suggested that large biochemical fluctuations can also be beneficial for organisms, ranging from bacteria to humans \cite{Aus06,Rao02,Has00,Pau00}. 
In a recent review article \cite{Los08}, Losick and Desplan have summarized a number of studies showing that certain cells choose one or another pathway of differentiation stochastically, without regard to environment or history.

Another example of stochastic signal processing is provided by the well-understood bacterial chemotaxis network. The flagellar motor of the bacterium is normally rotating in the counterclockwise (CCW) direction, but shows stochastic intervals of clockwise (CW) rotation. This gives rise to distinct phases of straight swimming motion of the bacterium, separated by random tumbling phases. Cell-membrane receptors detect the concentration of attractant molecules in the surrounding medium of the bacterium. Over several intermediate steps, the activation level of the receptors affects the distribution of CCW interval length and, thereby, the run length distribution of the bacterium's random walk in the medium. A statistical analysis of the CCW intervals revealed a power-law distribution \cite{Kor04}, which has been related to molecular noise in the reaction network \cite{Tu05}. Interestingly, such random walks with power law-distributed run lengths (Levy-flights) are known to generate trajectories which are the optimum strategies to search efficiently for randomly located objects \cite{Vis93}. This example shows how the shaping of molecular noise and the modulation of the noise parameters in response to environmental stimuli can be used by cells for complex tasks, such as foraging behavior.

We note that similar ideas of stochastic signal processing have recently emerged in the field of neuro science \cite{Gro09}. In the new concept of `reservoir computing', a network of (randomly) connected neurons generates a so-called transient state dynamics, where the trajectory of the system state is temporally fluctuating between various unstable attractors. This autonomously active `reservoir' network is only weakly coupled to the `input' and `output' units. As simulations have shown, the mapping of low dimensional input signals onto the high dimensional state space of the reservoir network can be advantageous for the signal processing. 

Finally, in this report we have discussed the stationary behavior of a single CMC in which the total number of molecules is fixed.  In living cells, however, multiple CMCs are connected in linear and branched signaling networks.  Moreover, the total number of molecules fluctuates as new proteins are expressed or old proteins are recycled.  If already a single CMC under stationary conditions gives rise to such highly complex, bizarre and non-deterministic behavior as described in this report, we argue that concentration fluctuations in living cells are even less predictable by classical mass action theory.





\section{\label{sec:app}Appendix}



\subsection{Average production rates}
\label{prod}
We consider an enzymatic conversion reaction of the general form:
\begin{equation}
X + E \autorightleftharpoons{$b$}{$u$} Y \autorightarrow{$c$}{} Z + E
\end{equation}
Using mass action rate theory, we obtain for the temporal change of the concentration y(t) of the enzyme-substrate complex:

\begin{equation}
\dot{y}= b\; x\; e - u\; y - c\; y
\end{equation}

We make the simplifying approximations that x(t) is held constant. After a certain relaxation time, the system will reach a steady state, in which also y(t)=const. The condition $ \dot{y}=0 $ then leads to

\begin{equation}
y = x \;e \frac{b}{u+c}.
\end{equation}

The expression $ \frac{b}{u+c} = \frac{1}{k_m} $ is defined as the inverse Michaelis constant, so that $ y=\frac{x\; e}{k_m} $. Since the enzyme can either be free or bound in the complex, $ e_t=e+y $, one obtains $ y=\frac{x (e_t-y)}{k_m} $. Solving for y yields 

\begin{equation}
y = e_t \frac{x}{x+k_m}.
\end{equation}

For the quantity of interest, the steady state generation rate $ \dot{z} = c\; y $ of the product, we finally obtain

\begin{equation}
\dot{z} = \left( c\; e_t \right) \frac{x}{x+k_m} = v_m \frac{x}{x+k_m}.
\end{equation}


\subsection{Enzymatic conversion as an effective Poisson process\label{poiss}}

In general, an individual $A$-enzyme molecule can undergo a series of binding/unbinding events with the (non-exhaustible) substrate $X_0$, before the substrate is finally converted into a new $X$-molecule. Therefore, even though each elementary reaction step, i.e. binding, unbinding and conversion, is a Poisson process, the same is not true for the multi-step production process. 
\footnote{For a simple example, consider a sequence of one binding and one conversion step. The PDF of each elementary Poisson step is exponential. The PDF of the sequence is a convolution of two exponential functions, i.e. a Gamma distribution with shape parameter $ k=2 $.}

However, many individual $A$-enzyme molecules, dispersed throughout the volume of the container, are simultaneously active, with independent temporal statistics. Our numerical simulations show that the superposition of many independent non-Poisson processes can resemble an effective Poisson process very closely. As expected, the characteristic time constant of this effective Poisson process is given by the inverse of the average total production rate $ \overline{R}_{act}(x_0) $.

In our CMC system, the substrate molecule number $ x_0 $, and therefore $ \overline{R}_{act}(x_0) $, are not constant. The resulting Poisson process is therefore not stationary but has a time-varying rate. 

We conclude that in systems with many independent enzyme molecules, the overall conversion process can be approximated by an inhomogeneous Poisson process.


\subsection{Poisson process as white Gaussian noise\label{gauss}}

Assume now a Poisson process with constant average event rate $ \overline{k}=\overline{R}_{act}(x_0)  $. We express the temporal change of the number x(t) of product molecules in the form

\begin{equation}
\dot{x}=\overline{k}+\Delta k(t).
\end{equation}

For later convenience, we want to approximate the fluctuation term by Gaussian white noise,

\begin{equation}
\left< \Delta k(t) \Delta k(t^{\prime})\right>=\Gamma \delta(t-t^{\prime}).
\label{correlator}
\end{equation}

What is the proper choice for the pre-factor $ \Gamma $, so that the major statistical properties of a Poisson process are consistently reproduced ?

To answer this question, we consider the number n(T) of X-molecules which are produced during an interval of length T:

\begin{equation}
n(T)=\int_0^T \dot{x}(t) dt = \overline{k}T +\int_0^T \Delta k(t) dt = \overline{n}+\Delta n.
\end{equation}

In the ensemble average, a Poisson process must fulfill

\begin{equation}
\left< (\Delta n)^2 \right> = \overline{n},
\end{equation}

or

\begin{equation}
\left< \left( \int_0^T \Delta k(t) dt\right)^2 \right> = \overline{k}T.
\end{equation}

The left side of the above equation can be reduced to $ \Gamma T $. Using Eq.(\ref{correlator}), we therefore obtain $ \Gamma = \overline{k} $, and therefore

\begin{equation}
\left< \Delta k(t) \Delta k(t^{\prime})\right>=\overline{k} \delta(t-t^{\prime}).
\end{equation}

Dividing this equation by $ \overline{k}$ leads to

\begin{equation}
\left< \frac{\Delta k(t)}{\sqrt{\overline{k}}} \frac{\Delta k(t^{\prime})}{\sqrt{\overline{k}}}\right> = \delta(t-t^{\prime}).
\end{equation}

We now define a new stochastic process by

\begin{equation}
\zeta(t)=\frac{\Delta k(t)}{\sqrt{\overline{k}}}.
\end{equation}

It is also normally distributed, but shows the desired property of $\delta$-autocorrelation with unit strength:.

\begin{equation}
\left< \zeta(t) \zeta(t^{\prime}) \right> = \delta(t-t^{\prime}).
\end{equation}

We conclude that a proper description of a Poisson process by a stochastic differential equation should have the form

\begin{equation}
\dot{x}=\overline{k} + \sqrt{\overline{k}} \; \zeta(t).
\end{equation}
















\begin{acknowledgments}
This work was supported by the ``Deutsche Forschungsgemeinschaft (DFG)''. We thank James Smith for stimulating discussions.
\end{acknowledgments}

\bibliography{apssamp}

\end{document}